\documentclass[a4paper,review,number]{elsarticle}
\usepackage{graphicx}
\usepackage{amssymb}
\usepackage[usenames]{color}

\begin{document}

\title{Coherent Electron Transfer in Polyacetylene}

\author[rvt]{D.~Psiachos\corref{cor1}\fnref{fn1}}
\ead{dpsi@physics.uoc.gr}
\cortext[cor1]{Corresponding author}
\fntext[fn1]{Tel.: +302810394246}
\address[rvt]{Crete Center for Quantum Complexity and Nanotechnology, Department of Physics, University of Crete, Heraklion 71003, Greece\\\vspace{3mm}\Large\textup{\textbf{CCQCN-2014-25}}}

\begin{abstract}
We examine, using mixed classical-quantum electron-ion dynamics, 
electron transfer in a donor-acceptor-like molecular junction system based on polyacetylene.
We identify two qualitatively-different transfer regimes: hopping and tunnelling. We discuss
the criteria for achieving each one and for minimizing inelastic scattering and 
decoherence arising from the coupling to the
ions, and we connect our main results to quantities derived from electron dynamics involving 
simpler, three-state model systems. We identify
the requirements to have near-ballistic transfer.
\end{abstract}

\maketitle

\section{Introduction}

Electron transport through organic molecules or chains of molecules
has been an active area of fundamental research because of its potential applications to molecular 
electronics. In Ref.~\cite{japan}, electron transfer in a system constituting a ``molecular
resonant tunnelling diode" device is studied and it is hypothesised that 
polaron levels form in the energy gap and that these facilitate
electron tunnelling across the molecular junction under certain voltage bias 
conditions. A simple expression for the electronic conductance through a molecule
is provided by the Landauer formula, which states that the conductance between two leads is proportional 
to the transmission. The behaviour of the transmission depends on the scattering processes 
involved. Some of the methods used to describe the scattering include Green's function methods, 
transfer matrix approaches such as the electron scattering quantum chemistry (ESQC) 
method~\cite{esqc}, or the use of
Lippmann-Schwinger scattering operators for mapping the non-equilibrium system onto an equilibrium 
one~\cite{han}. All of these
methods are time-independent, thus enabling the use of the Landauer or, in the case of several leads, the 
Landauer-B\"{u}ttiker formula, to evaluate the current. Non-equilibrium Green's 
function-based approaches or approaches based on the time-dependent Schr\"{o}dinger equation
applied to the key parts of the system~\cite{bishop,renaud,frederiksen}, enable a direct evaluation of the 
transmission and the current. 

The treatment of the electron-ion interaction within semi-empirical formalism assumes the validity of one of
two regimes: strongly or weakly-localised electron-ion interaction. The Marcus theory~\cite{marcus} treats
electron transfer non-adiabatically with a rate expression involving the initial, donor, and final, 
acceptor, states and geometries expressed
in terms of two key quantities, presumed to be separable owing to the slowness of the intermolecular vibrations
compared to the fast electron transfer rate: the difference in energy of the ground-state configurations 
before and after electron transfer, and the non-adiabatic contribution, called the reorganisation energy, which
is defined as the electronic excitation energy from the initial state in its ground-state configuration to the
electronic state of the product. As in
the Marcus assumption of distinct donors/acceptors, the Holstein model~\cite{holstein}
also treats the electron as being highly localised in space. In general, these techniques are most 
appropriate for non-homogeneous systems or systems containing defects or other strong scattering centres 
where there is a clear distinction between donor and acceptor and the transfer is accomplished by 
hopping between these two species. In more homogeneous systems,
there tends to be a more delocalised electron-ion interaction, leading to relatively adiabatic 
electron transfer. 

Different models for electron transfer
have been developed with the goal of describing polaron motion in oligomers. The Holstein model describes
``small-polaron" formation, or localised interaction with a particular molecule (site) and the electron conduction
is accomplished by distinct hopping, a non-adiabatic process. The structural perturbation is similarly highly localised. Another 
regime, the ``large-polaron" model, which leads to a longer-ranged structural distortion and at the same time,
a long-ranged electronic effect, is more suitable for periodic systems. The best-known description for 
large-polaron formation is the SSH~\cite{ssh-rmp,ssh-pnas,ssh-prb} model. Polaron transport in oligomer chains using the SSH model 
driven by an electric field has been studied extensively~\cite{conwell-apl,conwell-synthmet2,stafstrom-prl}. %The 
Different types of electron-phonon systems have also been studied
using fully-quantised models, which also include electron correlations~\cite{fehnske}. 

It should be noted that hopping and tunnelling modes of
transfer are not particular to the choice of model used (large or small-polaron) but depend also on the 
parameters used~\cite{tikhodeev}. Recently, the Holstein and SSH models have been combined, 
in order to describe both the inter- and intra-molecular interactions respectively of a 
crystal formed of pentacene chains~\cite{ishii}. While the inter-molecular interactions, originating
from Van der Waals forces, are weak, they are not neglected because in realistic molecular crystals there is a high
degree of disorder in the molecules' arrangements, and this was found to play a significant role
in determining the type of charge transport at nonzero temperatures - going from band-like to diffusive hopping transport
as the disorder increases. Traditionally, the distinction between the two types of transport has been made on
the basis of thermally rather than phonon-activated motion in polaron models~\cite{kalosakas}.

 In the present study, we work with a single chain of atoms only, within the ``large-polaron" regime. We
 study the influence of electron-ion
coupling on electron transfer in a polyacetylene (PA) chain coupled to molecules at both of its ends 
such that a donor-acceptor-like system is formed. Rather than using an electric field, we create 
time-dependent dynamics by
considering unbound electrons localised initially on an end-site molecule acting as a donor 
and shifting to acceptor status once the electron
reaches the end-molecule at the other side. We study the influence of the end-site molecules as well
as the coupling to the PA chain on the electron-ion dynamics timescales and the 
charge propagation through the chain.

\section{Methodology}
\label{sec:methodology}
\subsection{The SSH Hamiltonian}
We use the SSH Hamiltonian~\cite{ssh-prb} to describe the electronic and structural properties of the PA chain. This model
uses a basis of site-centred $\pi$ orbitals in the tight-binding approximation to describe the electronic properties,
while the ions are treated classically. Regardless of the number of electrons, the basis size is fixed 
to the number of atomic sites, as per the LCAO approximation. Explicitly, the 
Hamiltonian is split into electronic and ionic parts:
\begin{equation}
\hat{H}=\hat{H}_{el}+\hat{H}_{ion}
\end{equation}
where
\begin{eqnarray}
\hat{H}_{el}=&-&\sum_{n,\sigma} t_{n+1,n}\left(c_{n+1,\sigma}^\dagger c_{n,\sigma}+c_{n,\sigma}^\dagger c_{n+1,\sigma}\right)\nonumber\\
&+&\sum_{n,\sigma}\epsilon_nc_{n,\sigma}^\dagger c_{n,\sigma}
\label{Hel}
\end{eqnarray}
is the electronic part where
the operator $c_{n,\sigma}^\dagger$ adds an electron of spin $\sigma$ 
to site $n$ and where in the SSH model, the site energies $\epsilon_n$ are set to zero and 
\begin{equation}
t_{n+1,n}=t_0-\alpha\left(u_{n+1}-u_n\right)
\label{tu}
\end{equation} describes the hopping of the electron. The hopping is 
modified from the fixed value $t_0$ by the electron-phonon coupling parameter $\alpha$ according to the displacements $u_n$ of the ions, from their
positions in a perfect, undimerised chain. The hopping range is confined to nearest neighbours. 

The ionic part,
\begin{equation}
\hat{H}_{ion}=\hat{V}_{ion}+\hat{T}_{ion}=\frac{1}{2}\sum_n K\left(\hat{u}_{n+1}-\hat{u}_n\right)^2+\sum_n \frac{\hat{p}_n^2}{2M}
\end{equation}
consists of a harmonic potential energy operator with spring constant $K$ and a kinetic energy operator. 
The set of parameters used is standard in the SSH model literature as applied to PA chains~\cite{ssh-prb}:
$t_0$=2.5 eV, $\alpha$=4.1 eV/\AA, $K$=21 eV/\AA$^2$, and $M$=1349.14 eV~fs$^2$/\AA$^2$. The electronic 
and ionic Hamiltonians 
are thus linked by the ionic displacements $u_n$.

\begin{figure}[htb]
\includegraphics[width=9.cm]{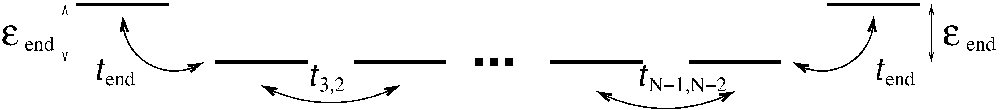}
\caption{Chain with nearest-neighbour hoppings varyinh according to the site 
distortions (see Eq.~\ref{tu}), coupled to molecular levels at its ends. The
total number of sites including the molecules at the ends is $N$.}
\label{chain}
\end{figure}

We include the site-energy term in Eq.~\ref{Hel} because we will consider a chain where each end is coupled to an end-site molecule
with energy level $\epsilon_{end}$. All other $\epsilon_n$ are set to zero. In addition, 
we will modify the coupling between the end-molecule level and the chain to a value $t_{end}$.
We will take the molecular levels and couplings at both ends to be the same. A schematic of a chain with its end sites coupled 
to molecular levels is shown in Fig.~\ref{chain}. The coupling $t_{end}$ is fixed and does not depend on the distortion of the chain's end sites.

\subsection{Solving the coupled ion-electron system}
We use the Ehrenfest approximation~\cite{tully,mele,streitwolf,allen,stafstrom-prl}, a non-adiabatic semi-classical 
scheme in which the ions undergo classical motion
while the electrons are treated quantum-mechanically, evolving in time. The adiabatic approximation, also known as 
Born-Oppenheimer (BO) dynamics, a simpler scheme in 
which the electron orbitals are treated time-independently, affected only by changes in the ionic potential, and 
thus not deviating from their initial superpositions of Hamiltonian
eigenstates, would not lead to electron transfer which is what we want to study here.
To that end, the equation of motion of the electrons $\left|\psi_k(t)\rangle\right.$ of state $k$ is given by
\begin{equation}
i\hbar\frac{d\left|\psi_k(t)\rangle\right.}{dt}=\hat{H}_{el}\left|\psi_k(t)\rangle\right.
\label{tdse}
\end{equation}
while that for the $nth$ ion is 
\begin{eqnarray}
M\frac{d^2u_n}{dt^2}&\equiv& F_n=-\left\langle\frac{d}{du_n}\hat{H}\right\rangle\nonumber\\
&=&-\sum_{k,occ}\left\langle\psi_k(t)\left|\frac{d}{du_n}\hat{H}_{el}\right|\psi_k(t)\right\rangle-\frac{d}{du_n}V_{ion},
\label{eom_ion}
\end{eqnarray}
which we integrate using the velocity-Verlet algorithm
\begin{equation}
u_n(t+\tau)=\frac{\tau^2}{M}F_n+2u_n(t)-u_n(t-\tau)
\end{equation}
for a sufficiently-small timestep $\tau$ with which to maintain energy conservation - \textit{e.g.} a $\tau$ of 0.02 fs conserves
the total energy to within 0.13 meV/site over 1 ps. In Eq.~\ref{eom_ion}, we use the definition of
the Hellman-Feynman force because it is this and not the derivative of the negative of the electronic energy
 which yields the
expectation values of the ionic trajectories. In practice however,
we found virtually no difference between the results obtained by the two force expressions for systems as large as the ones we studied. 

Letting the electronic eigenstates arising from a solution to the time-independent Schr\"{o}dinger equation
\begin{equation}
\hat{H}_{el}\left|\chi_m\rangle\right.=\epsilon_m\left|\chi_m\rangle\right.
\end{equation}
where $\hat{H}_{el}$ implicitly depends on the ionic configuration $\{u_n\}$,
 be represented by the set $\{\chi_m\}$, 
the wavefunctions of all the electrons $k$ can be written as superpositions of the eigenstates
as 
\begin{equation}
\left|\psi_{k}(t)\rangle\right.=\sum_m g_{m,k}(t)\left|\chi_m\rangle\right.
\end{equation}

The initial conditions of the problem are divided conceptually into two types: the first type of initial condition is $g_{m,k}(0)=\delta_{mk}$ 
and is imposed on the $N$ wavefunctions
of the $N$-site system, while the extra electron is set up in a superposition state in general. Thereafter,
all of the $\psi_{k}(t)$ evolve according to Eq.~\ref{tdse} along with the boundary condition that they are continuous across time steps, and become non-stationary states, written in terms of the eigenstates corresponding 
to the extant value of $\{u_n\}$.
  
In all of the results presented in this paper, we treat all of the electronic wavefunctions with the Ehrenfest 
method. Before an extra electron is added to the system, the ground state of the neutral chain, whether coupled 
to a junction or not, is determined. This is
done via the procedure described above with the ionic velocities held to zero. We take the initial conditions
of the electron-ion dynamics problem to correspond to zero temperature. 
Here, the ends, which may correspond to end-site molecules, are free to move while a total length constraint is imposed on the system. The purpose of 
this is to obtain displacement values for the ends which are consistent with the parameters
of the Hamiltonian, thereby minimising the finite-length effects. Afterwards, for the 
dynamics, the length constraint is removed and the end states are held fixed to their relaxed values because keeping the length constrained
instead of fixing the ends led to the plots of displacement $u$ versus site number being tilted, as the chain would translate, and
this made the results difficult to 
compare and analyze. Imposing no boundary conditions whatsoever results in highly-distorted 
ground-state structures due to the finite length of the chains, obscuring the phenomena which we would like to study.

\subsection{Semi-classical approximation}
The classical treatment of the ions within the SSH model has been questioned 
in the past~\cite{McKenzie,Su} since
is well-known that the higher-frequency $\hbar\omega\gg k_BT$ vibrational 
modes are not well-treated by classical mechanics. In Ref.~\cite{franco},
 the authors use classical mechanics for describing the ion motion 
but consider that the lowest vibrational mode of the system imparts 
kinetic energy to the ions at zero temperature. It is worth precluding
the effect of zero-point amplitude fluctuations on the results of our 
calculation. Ness and Fisher~\cite{nessfisherPRL} found that, between
two models, one having quantised ionic vibrations and the other 
classical, there was almost no difference in the 
dimerisation pattern for a 40-site chain containing a polaron. 

Since we find that 
our system follows a Debye-model density of states, the dynamics at 
low temperatures is dominated by the lowest normal 
mode~\cite{yangkawazoe}. It is valid to ignore the contribution 
of the zero-point energy (ZPE) $\hbar\omega_0$ for long-enough 
chains such that $\hbar\omega_0\leq k_BT$. ZPE
here refers to the lowest normal mode of the total system, rather
than the lowest modes of the constituent atoms. In this study, we focused 
on chains of length 60 sites. For these, the $\hbar\omega_0$ of 
27~meV is about equal to $k_BT$ (25 meV) at room temperature 
so the classical approximation should be 
accurate. Examining longer chains, 80 sites gives $\hbar\omega_0$=20 meV, 
while for 100 sites the value is 16 meV so our neglect of the ZPE is 
expected to become more accurate as longer chains are considered. We 
have repeated one of our cases of ballistic transfer 
(Fig.~\ref{e0.7t-0.4} later in Sec.~\ref{Sec:examples}) with an 80-site
 chain and we find the same type of behaviour in general 
terms as for 60 sites. 

Franco and Brumer~\cite{franco} have 
performed their electronic coherence dynamics calculations on 
short (\textit{e.g.} $<$15 sites) and long (50, 100 sites) 
chains and find that due to the dense electronic spectrum of the 
latter, decoherence occurs exclusively through nearby electronic levels and
nuclear overlaps do not play a role. In contrast, nuclear overlaps 
arising from anharmonic excited-state potential energy surfaces are the 
main mechanism of decoherence in short chains and a careful examination 
of ionic initial states is very important for these. Further, in the
case of short chains, a quantum treatment of phonons
could even be required throughout the dynamics, while for long
chains, the classical description is sufficient.

\section{Systems studied}
\subsection{Isolated PA chain}
The system corresponding to an isolated PA chain is a special case of Eq.~\ref{Hel}, in which the end-site 
molecules are given an energy $\epsilon_{end}=0$
and coupling $t_{end}$ according to Eq.~\ref{tu} \textit{i.e.} the coupling was not fixed.
The electron-ion dynamical behaviour for such a system of $N$ sites and $N$ non-interacting 
electrons is well-known in the literature. The ground state of the even chain
leads to the well-known dimerisation pattern, shown in Fig.~\ref{u}a, while that of the odd chain 
contains a soliton, both as expected~\cite{ssh-prb}. 

\begin{figure}[htb]
\includegraphics[width=8.5cm]{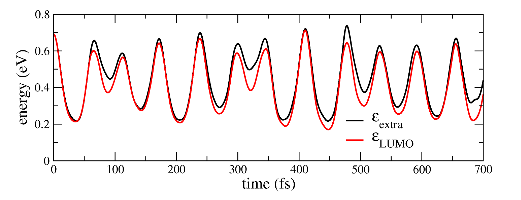}
\caption{The energy of the added electron deviates from the LUMO, in Ehrenfest dynamics.}
\label{polaron}
\end{figure}

After obtaining these ground states, an electron was prepared in either the LUMO, or in a non-stationary state at one end, 
and the dynamics were performed. In the even-site chain, the case of the LUMO preparation 
leads to a polaron~\cite{ssh-prb,conwell-synthmet2,phillpot}, a structural
deformation consisting of a soliton/anti-soliton pair moving in tandem with the electron localisation. We focused on even-site
chains in order to avoid considering soliton dynamics along with the transfer of the extra electron. The effect of the Ehrenfest
approximation as opposed to the BO one is clear from Fig~\ref{polaron}. The energy of the extra wavepacket, initially prepared in the 
LUMO, closely follows the LUMO but nonetheless deviates from it at times. 

\begin{figure}[htb]
\includegraphics[width=8.5cm]{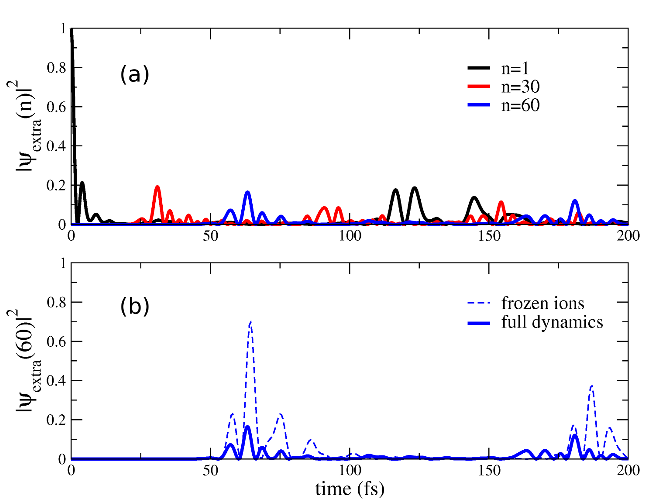}
\caption{(a) The populations of the two end sites 1, 60, and a central site 30, for electron-ion dynamics for a 
chain not coupled to molecular junctions
where an extra electron was added at site 1 at t=0. (b) Comparison with the frozen ion case. The population shown is only
due to the extra electron (see text for an explanation).}
\label{resonance}
\end{figure}

\begin{figure}[htb]
\includegraphics[width=8.5cm]{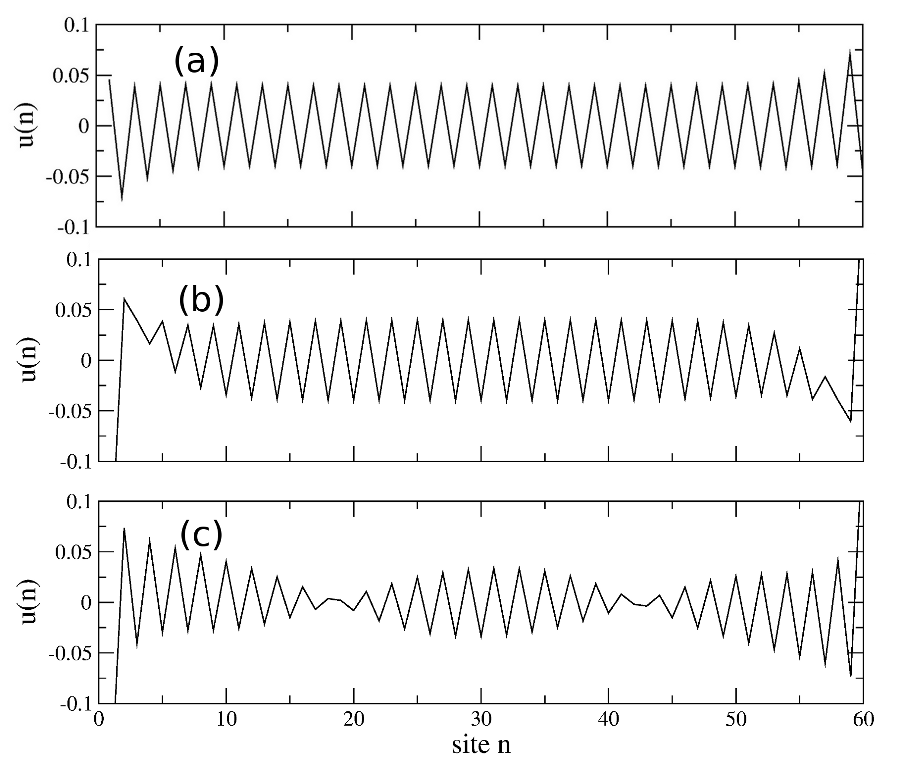}
\caption{(a) Ionic distortions $u(n)$ in~\AA for a chain with no coupling to a molecular junction at its ends, (b) $\epsilon_{end}$=0.3 eV and $t_{end}$=1.0 eV,
and (c) $\epsilon_{end}$=0.8 eV and $t_{end}$=0.4 eV.}
\label{u}
\end{figure}

For the electron prepared at one end, the charge density oscillated back and forth (see Fig.~\ref{resonance}a). This 
oscillation was virtually
all due to the extra electron - the other electrons in the chain had practically no contribution to it. However, the
chain distortions were very large, several times greater than those corresponding to the dimerisation ($\sim\pm0.04$~\AA,
see Fig.~\ref{u}a), and 
in practice, would require anharmonic potentials to treat, and may lead to
chain breakup. A more advanced semi-classical treatment of ionic dynamics is presented for example in 
Ref.~\cite{Lu}. It is important to note that as the electron 
wavefunction oscillates back and forth, it distinctly passes through all sites between the two ends. Therefore, 
this is a hopping transfer regime as opposed to a tunnelling one. 

In order to 
determine the effect of the ions on the
electron propagation, we compare the propagation with that for the case of frozen ions 
(Fig.~\ref{resonance}b). In the frozen-ion case, the propagation
is quasi-periodic, owing to the non-commensurability of the eigenvalues in the eigenstate superposition comprising the wavefunction of the propagating
electron. For increasingly longer times, according to the Poincar\'{e} theorem applied
to quantum superpositions of eigenstates with non-commensurate eigenvalues, the oscillations will approach the original
value increasingly faithfully and eventually lead to a value near 1. This is the reason why the value of 1 is not reached during the 
few ps of the simulations for the frozen ion case. Beyond 
that, the electron-ion interaction destroys the coherences (off-diagonal entries of the density matrix), which manifests itself in 
a decrease in site populations in Fig.~\ref{resonance}b for the simulation with full dynamics, as compared with the frozen-ion case. The 
off-diagonal entries of the density matrix appear in the electron density
due to the extra electron as a result of intereferences from the eigenstate superposition.

\subsection{Chain within molecular junctions}
\label{Sec:examples}
The molecular junction coupling parameters $t_{end}$ and $\epsilon_{end}$ were varied across a wide range of values in order to search for
an optimal set which would lead to low electron decoherence, meaning a small influence of the ions on the electron transfer.
This required first obtaining a relaxed ground-state structure for the ions, and then performing dynamics using the added
wavepacket. We only performed dynamics for
cases which looked promising based on the frozen-ion transfer calculations. 
The ground-state structure for a chain coupled to a molecule at each end deviates
significantly from that of a dimerised chain (Fig.~\ref{u}a). Most notable, apart from the end effects, are the kinks introduced
at large $\epsilon_{end}$ and small $t_{end}$, (Fig.~\ref{u}c) which migrate towards the ends for smaller 
$\epsilon_{end}$ and larger $t_{end}$ (Fig.~\ref{u}b). 

\begin{figure}[htb]
\includegraphics[width=8.5cm]{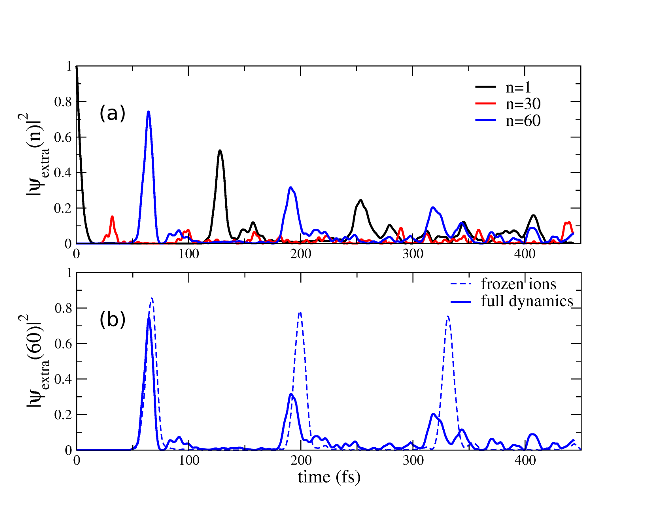}
\caption{(a) The populations of the two end sites 1, 60 and a central site, 30, for electron-ion dynamics for a 
chain coupled to end molecules with energy $\epsilon_{end}$=2.0 eV and coupling to chain $t_{end}$=1.1 eV
where an extra electron was added at site 1 at t=0. (b) Comparison with the frozen-ion case. Only the contribution
from the extra electron is shown (see text for an explanation).}
\label{e2.0t-1.1}
\end{figure}

Regarding the charge density oscillations, as with the case of the chain not coupled to any junction, only the extra electron reached the 
other side. The other electrons nevertheless caused a significant amount of 
charge-density oscillation, which was correlated with the time-evolution of the ionic deformation. In 
Fig.~\ref{e2.0t-1.1} we show a case also in the ``hopping" regime, albeit not as extreme as the 
case of Fig.~\ref{resonance} which had equal occupation of all sites. Still, there 
is a non-negligible occupation of all of the interior chain sites (only that of 
a central site is shown in the figures), 
and as discussed below, numerous Rabi frequencies, which do not
show up at the small times in which relatively-coherent oscillation is maintained. 

\begin{figure}[htb]
\includegraphics[width=8.5cm]{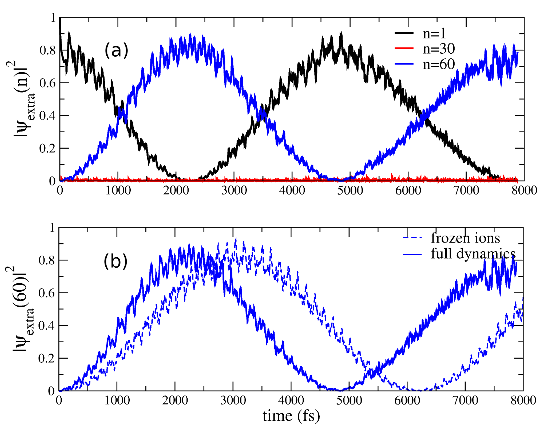}
\caption{(a) The populations of the two end sites 1, 60 and a central site, 30, for electron-ion dynamics for a 
chain coupled to end molecules with energy $\epsilon_{end}$=0.6 eV and coupling $t_{end}$=0.4 eV
where an extra electron was added at site 1 at t=0. (b) Comparison with the frozen-ion case. Only the contribution
from the extra electron is shown (see text for an explanation).}
\label{e0.6t-0.4}
\end{figure}
\begin{figure}[htb]
\includegraphics[width=8.5cm]{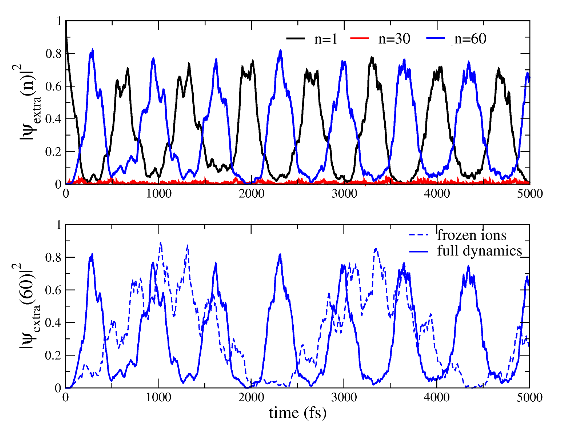}
\caption{(a) The populations of the two end sites 1, 60 and a central site, 30, for electron-ion dynamics for a 
chain coupled to end molecules with energy $\epsilon_{end}$=0.7 eV and coupling $t_{end}$=0.4 eV
where an extra electron was added at site 1 at t=0. (b) Comparison with the frozen-ion case. Only the contribution
from the extra electron is shown (see text for an explanation).}
\label{e0.7t-0.4}
\end{figure}

It is important to note
that distinct hopping across the chain by the extra electron was not always found to occur - we also identified another limit:
that of tunnelling. In Fig.~\ref{e0.6t-0.4} we show a case in the ``tunnelling" regime which had relatively 
low scattering of the electron by the ions. It is called tunnelling because there is almost no occupation of the chain sites,
with one dominant electron oscillation frequency connecting the defect levels arising from the end states (see Sec.~\ref{Sec:rabi}).

An interesting case, where the electron oscillation frequency changes dramatically with the dynamics, is shown 
in Fig.~\ref{e0.7t-0.4}. Also, the oscillations
are incredibly persistent over very long times. Generally, the electron oscillations are stable over
long times throughout the tunnelling regime 
because the electron tunnels between the two
fixed end molecules with minimal impact from any structural distortions which may be occurring in the chain.

\subsection{Rabi frequencies}
\label{Sec:rabi}
We systematically studied the oscillation of an electron in a frozen-ion configuration by varying 
the electronic parameters $t_{end}$ and $\epsilon_{end}$ of the end molecules.

Usually, the site population over time can be approximately represented
by a sinusoidal oscillation with a single (Rabi) frequency, which is defined as 
the frequency of oscillation for an electron in a three-state quantum system. We 
determined the contribution of the pair of eigenstates comprising each Rabi frequency to the
electron oscillation by projecting the eigenstates onto the end-molecule 
wavefunctions~\cite{joachim2}. Explicitly, this weight
for the (n,m) pair of eigenstates is equal to
\begin{equation}
D_{nm}=\left\langle\phi_1\right.\left|\hat{P}_n\right|\left.\phi_{end}\right\rangle\left\langle\phi_{end}\right.\left|\hat{P}_m\right|\left.\phi_{1}\right.\rangle
\label{D}
\end{equation}
where the operator $\hat{P}_i=\left|\chi_i\rangle\langle\chi_i\right|$ is the projector of eigenstate $i$
and the two end-molecule sites are labelled $\phi_1$ and $\phi_{end}$. In 
three-state quantum systems there is a relationship of inverse proportionality
between the Rabi frequency and the weight (\textit{viz.} the quantity $V$ in 
Eq.~3 in Ref.~\cite{joachim1}) which we exploit in 
our analysis of Sec.~\ref{Sec:model}. 

 Characteristics of the frozen-ion system which caused the electron transfer to be
relatively impervious to ionic vibrations were generally short Rabi oscillation timescales and
a single dominant Rabi frequency. At the same time, under such conditions, the reverse was 
also found to be true: the electron oscillation did not affect the structure much, which is a different 
result than the one found by Ness and Fisher~\cite{nessfisherPRL} who
used a model with a small number of quantised phonon modes. On the other hand, the presence of noise which
competed strongly with the main Rabi frequency led to the electron transfer being 
incoherent as a result of destructive interferences from the ions' effect on the 
different frequencies present (Sec.~\ref{Sec:model}).

The hopping case (Fig.~\ref{resonance}, and \ref{e2.0t-1.1}) is characterised by the presence of a large number of
competing Rabi frequencies while in tunnelling (\textit{e.g.} Fig.~\ref{e0.6t-0.4}), just one, or possibly 
two, frequencies dominate the 
oscillations. However, even in hopping, at small times there seems to be just one main frequency present, corresponding to
a fast time scale, on which there is superimposed noise from an even faster one. The other 
frequencies, with a longer period, eventually show up, and lead to an indistinct oscillation but by that time, the oscillation
is already destroyed by the electron-ion interaction. In Sec.~\ref{Sec:analysis} we discuss which choices of parameters
lead to these different regimes and how the transmission occurs.

\section{Analysis}
\subsection{Defect states}
\label{Sec:analysis}
\begin{figure}[htb]
\includegraphics[width=8.5cm]{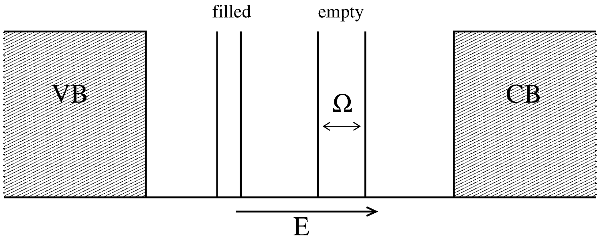}
\caption{Schematic of the energy levels for a system with two kinks in the distortion $u(n)$, \textit{e.g.} Fig.~\ref{u}c. In 
a case such as Fig.~\ref{u}b, there is no longer a dominant Rabi frequency $\Omega$ and transmission takes place through many levels,
including those in the conduction band (CB), filled defect levels, and valence band (VB).}
\label{e-levels}
\end{figure}

The perfect dimerised chain has a double bond at both ends and an energy gap of 1.39 eV. When $t_{end}$ is
significantly different from $t_0$, defect states appear in the gap. If the chain is still well-dimerised, as 
for the case of Fig.~\ref{u}b, 
there are only two defect states, one filled and one empty, degenerate and at nearly the energy $\epsilon_{end}$; 
otherwise there are two additional defect states, split off from the valence and conduction band respectively, consistent with 
a bi-polaronic distortion such as that shown in Fig.~\ref{u}c. The latter case is the most usual, as the case of only two defect states
is confined to marginal cases (very small $\epsilon_{end}$). Fig.~\ref{e-levels} shows the one-electron energy level 
structure of the system attached to end molecules, for the most common case, that of a bi-polaronic distortion.
 For low  $t_{end}$ the empty-state
defect levels are nearly degenerate and approximately equal to $\epsilon_{end}$. The bi-polaronic distortion is gradually lost for
$t_{end}>\epsilon_{end}$ and progresses to the ends for $t_{end}\gg \epsilon_{end}$, and for sufficiently large 
$t_{end}/ \epsilon_{end}$, there are no defect states at all in the gap.

The closeness in energy of the empty states for low $t_{end}$ means that
the Rabi oscillation period is very long. This is also the pure ``tunnelling" regime because there 
are virtually no other contributions from other pairs of eigenstates to the electron oscillation
between the ends. 

For $t_{end}\ll \epsilon_{end}$, the 
defect levels are present, owing to the bi-polaronic distortion through the system, but only the filled ones are in the gap while the empty ones are located 
in the CB, assuming that $\epsilon_{end}$ is large enough that it is in the CB. A very low $t_{end}$ combined with a 
large $\epsilon_{end}$ leads to a very small Rabi 
frequency, or very slow electron oscillation, which may be undesirable. Nevertheless, this is the pure
tunnelling regime, as these defect energy levels are highly localised at the end molecules. However, as $t_{end}$ increases,
more CB states around the energy $\epsilon_{end}$ suddenly become involved in the transfer leading
to a noisy spectrum. Thus, a large $\epsilon_{end}$ which leads to defect states in the CB is not very practical for 
achieving stable coherent oscillations. On the other hand, for $t_{end}\gg \epsilon_{end}$, there
are no defect levels as the chain is dimerised and the transfer occurs through a large number of levels,
both filled and empty. The extreme case is that of Fig.~\ref{resonance}. For both of these situations, the system
is in the hopping regime as the tunnelling through numerous pairs of energy levels implies that the wavefunction of the 
extra electron has a non-negligble projection onto
the chain sites. However, for intermediate values of $t_{end}$ and
$\epsilon_{end}$, we are able to obtain tunnelling and a fairly large, single dominant oscillation frequency, which is stable
under the effect of ion dynamics for at least a few cycles, over a wide range of molecular-junction parameters $t_{end}$ and
$\epsilon_{end}$.

\subsection{Effect of dynamics on Rabi frequencies}
\label{Sec:model}
Here, we describe the effect of ion dynamics on the Rabi frequencies, in order to explain the decoherence mechanism. A recent
study~\cite{franco} examined the decoherence mechanism for electron wavepackets added to chains and determined
that for long chains, decoherence is caused by population transfer into other states due to the dense energy level 
structure. Our results, which show a decay in population of the initial wavefunction
superposition, especially in the case of the defect levels located in the conduction band, agree with this explanation. 

\begin{figure}[htb]
\includegraphics[width=8.5cm]{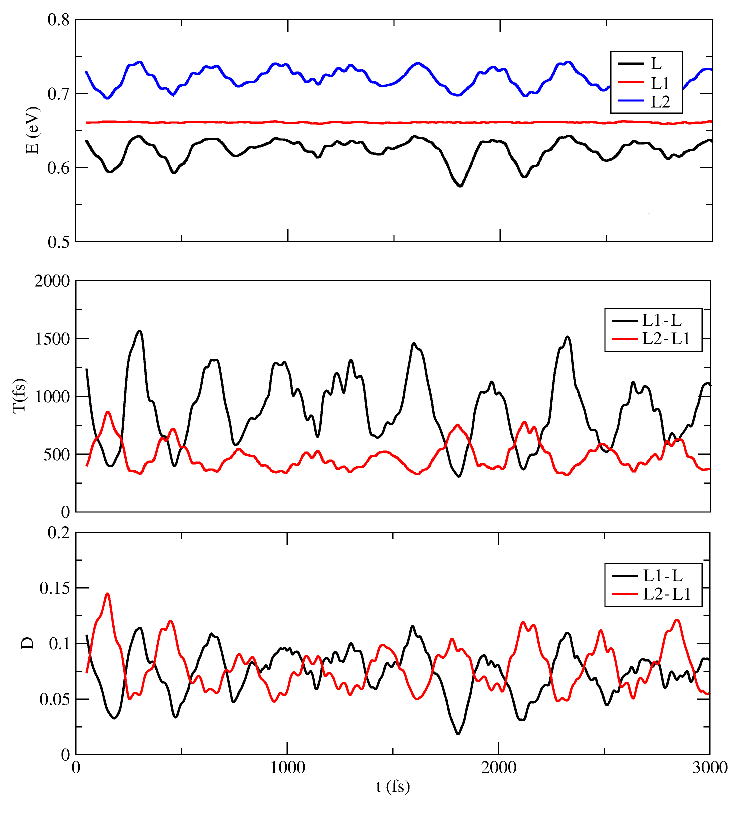}
\caption{Variation of key quantities with time for the case $t_{end}$=0.4 eV and 
$\epsilon_{end}$=0.7 eV. The top panel shows the eigenvalues, L: LUMO, L1: LUMO+1, L2: LUMO+2.
The middle panel shows the inverse of the Rabi frequency, $T(t)\equiv 1/\Omega(t)$ for the two competing
Rabi frequencies formed by the two pairs of energy levels shown. The bottom panel shows the variation with
time of the weights of these Rabi frequencies. For clarity, the data shown is a running average over 100 fs.}
\label{lnm-eig}
\end{figure}

In addition, we find that when multiple Rabi frequencies compete, they 
interfere. This is the situation in the hopping regime, and the coherency of the 
electron oscillation is substantially degraded
by the ionic dynamics, as also mentioned earlier in Sec.~\ref{Sec:analysis}. However, for several cases
of two competing Rabi frequencies, we find that the electron dynamics exhibit exceptional stability, particularly when
the two Rabi frequencies are close and contribute roughly equally to the oscillation. The reason is that their weights oscillate out of phase
with each other. In Fig.~\ref{lnm-eig} we show the time dependence of the energy levels 
involved in the oscillation, 
the inverse Rabi frequencies $T(t)$ derived from these energy levels, and the weights $D(t)$
of the latter, for a case which had great stability in the electron oscillation. As 
mentioned earlier in Sec.~\ref{Sec:rabi}, 
the quantities $T(t)$ and $D(t)$ are correlated. We create a simple model
for combining the two Rabi oscillations, omitting the time dependence from $T(t)$ as it is already found in $D(t)$.
If we denote the average value of the $i$th inverse Rabi 
frequency (connecting two particular energy levels) as $\overline{T}_i\equiv 1/\overline{\Omega}_i$,
and its weight $D_i(t)$ (see Eq.~\ref{D}), composed of a mean value $\overline{D}_i$ and a term 
of amplitude $B_i$ oscillating with a frequency $\omega_0$, we may write
\begin{eqnarray}
T_1(t)&=&\overline{T}_1\nonumber\\
D_1(t)&=&\overline{D}_1+B_1cos(\omega_0t)\nonumber\\
T_2(t)&=&\overline{T}_2\nonumber\\
D_2(t)&=&\overline{D}_2-B_2cos(\omega_0t)
\label{TD}
\end{eqnarray}
and the combined oscillation period becomes:
\begin{equation}
T_{net}(t)=\frac{\overline{T}_1{D}_1(t)+\overline{T}_2{D}_2(t)}{D_1(t)+D_2(t)}.
\label{Tnet}
\end{equation}

We found that this expression represents the electron oscillation period very well. The main cause
of any deviation from the prediction of Eq.~\ref{Tnet} and the data is due to
other secondary inverse Rabi frequencies, which are smaller and tend to bring the actual value down. The 
time-dependence in Eq.~\ref{Tnet} is generally quite weak, owing to the opposite signs
in the time-dependent quantities of Eq.~\ref{TD} and this leads to the exceptional
stability seen in Fig.~\ref{e0.7t-0.4}. We determined the average value $T_{av}$ of $T_{net}(t)$ and
compared it with the main oscillation period for several cases, which we show in Table~\ref{rabitab}. Regarding
the oscillation frequencies $\omega_0$, they mainly corresponded to the bare (without the extra electron) lowest
phonon modes of the respective system but not always, and importantly, no relation was seen between the matching or not
of the $\omega_0$ with a bare phonon frequency and the stability of electron oscillations in the respective system.
\begin{table*}[ht]\footnotesize
\begin{tabular*}{1.07\textwidth}{l| l l l l l l l l}
$\epsilon_{end}(eV)$&T(0)&$\overline{T}_i$&$\overline{D}_i$&$B_i$&$2\pi/\omega_0$&T$_{av}$&T$_{actual}$\\
\hline
0.6&6239&4286&0.116&0.012&172, 300, 250&4286&3879\\
0.62&5274&3218&0.132&0.011&211&3218&3100\\
0.64&4331&2229&0.119&0.011&269&2229&1900\\
0.65&3900&2014&0.113&0.010&303, 160&2014&1520\\
0.67&3158&1004 (483)& 0.073 (0.069)&0.007 (0.006)&443&751&740\\
0.7&2231&882 (495)&0.072 (0.081)&0.025 (0.025)&328&678&650\\
0.72&1681&863 (521)&0.073 (0.090)&0.036 (0.034)&278&674&574\\
0.75&1110&639 (593)&0.056 (0.106)&0.016 (0.036)&250&609&615\\
0.77&853&513 (611)&0.044 (0.115)&0.009 (0.025)&216&584&640\\
0.8&586&672&0.120&0.022&169&672&586\\
1.0&440&505&0.114&0.023&125&505&469
\end{tabular*}
\caption{Parameters of Eq.~\ref{TD}-\ref{Tnet} for several cases of $\epsilon_{end}$ (in eV) for 
fixed $t_{end}$=0.4 eV, as deduced from the dynamics, for comparison with the average value
T$_{av}\equiv \overline{T_{net}}(t)$ calculated from Eq.~\ref{Tnet} and with the main
electron oscillation period $T_{actual}$ observed from the simulations. Note 
$T(0)$, the inverse frequency corresponding to having frozen ions, is given for 
comparison. If two inverse Rabi frequencies
compete, the values corresponding to both are given, the same one always in parentheses. The
units for time are fs.}
\label{rabitab}
\end{table*}
 
All of the electron oscillations speed up when the ionic dynamics are included in the calculation and
the explanation is provided by the behaviour of the relevant eigenvalues and inverse Rabi frequencies from Fig.~\ref{lnm-eig}.

\section{Conclusions}

In summary, we have investigated the inelastic electron transfer in a polyacetylene wire within molecular junctions and we found two
distinct transfer regimes: tunnelling and hopping. Furthermore, under certain conditions, we find near-ballistic 
tunnelling transfer caused by the interference of two dominant Rabi frequencies. The tunnelling case 
resulted when the defect levels from the electrodes were in the energy gap of the chain. When
the defect levels are in the conduction or valence band, tunnelling occurs through multiple energy levels leading to a non-negligible
occupation of the chain sites, a regime which we term hopping but as we are working at zero temperature this term is not to
be confused with thermally-activated hopping~\cite{kalosakas}. 

Ness and Fisher~\cite{nessfisherPRL} have calculated tunnelling electron transport for a chain coupled to metallic leads, using 
a model incorporating a quantised description of the main phonon modes, and find a clear indication of polaron 
transport, or a distortion of the ions which propagates along with the electron. On 
the other hand, another study~\cite{bishop},
based on the SSH model with a classical-oscillator description of ionic vibrations, does not find any polaron transport. It does
find however, that energy levels arising from a pre-existing polaronic distortion aid in the tunnelling process. Our work suggests that not 
polaronic \textit{per se}, but defect levels in the energy gap, produce tunnelling transfer. 

We expect that there 
will be dramatic changes in the electron transfer characteristics, particularly
for the cases in which the defect levels are located in the conduction band, if we consider electron-electron interactions,
and this extension is presently under investigation. Also presently under investigation is an extension towards a
system where the end-site molecules in the junction are replaced by electrodes. It is not expected that the essential results
of this study will change upon adding electrodes, whose disspiative effect would be to broaden the energy-level structure. Complicating
matters is the influence of the geometry of the electrodes~\cite{geometry1,geometry2}. 

Other future extensions could include a
quantum-mechanical treatment of the ions~\cite{fehnske}, and a consideration of initial conditions
corresponding to non-zero temperature. Even though quantum effects 
of the lowest mode can safely be neglected for long chains, the model
could be made more realistic if an initial distribution of ionic kinetic energies
consistent with the temperature at which the neglect of the ZP fluctuations is 
strictly valid were used. Still, our main finding of ballistic transfer, or electron transfer unaffected by the ionic vibrations in the chain, would not be expected to change if the ions were given an initial thermal distribution.     

While we find that multiple competing Rabi frequencies lead to hopping and a destruction of the 
electron oscillations because of their different behaviour under the dynamics, we find a large stable 
regime where two competing Rabi frequencies interfere destructively in order to yield exceptionally-stable electron
oscillations. 

The case of pure tunnelling (small $\Omega$) is similar to the case of electron transfer
across a junction by resonant tunnelling described in a recent experiment~\cite{japan}. The 
results of the present work correspond quite closely with the schematic in Fig.~5b or c 
of that study. In 
the present work, the ends correspond to the PDA chains of the experimental study, while the 
polyacetylene chain corresponds to the inner ZnPc molecule of that study. The ends in the present work 
are unbiased, and at the same energy, but they
introduce polaronic defect levels (two filled and two empty, as in Fig.~5 of 
Ref.~\cite{japan}) shifted slightly in energy by $\Omega$ due to the mixing with the states
of the inner molecule. The tunnelling 
between the ends may occur ballistically - without scattering by the inner
molecule, with the choice of certain end-site coupling parameters. 

\section*{Acknowledgements}
This work was partially supported by the European Union's Seventh Framework 
Programme (FP7-REGPOT-2012-2013-1) under grant agreement 316165.


\begin{thebibliography}{10}
\expandafter\ifx\csname urlstyle\endcsname\relax
  \providecommand{\doi}[1]{doi:\discretionary{}{}{}#1}\else
  \providecommand{\doi}{doi:\discretionary{}{}{}\begingroup
  \urlstyle{rm}\Url}\fi

\bibitem{japan}
Y.~Okawa, S.~K. Mandal, C.~Hu, Y.~Tateyama, S.~Goedecker, S.~Tsukamoto,
  T.~Hasegawa, J.~K. Gimzewski, M.~Aono,
\newblock J. Am. Chem. Soc. 133 (2011) 8227.

\bibitem{esqc}
P.~Sautet, C.~Joachim,
\newblock Phys. Rev. B 38 (1988) 12238.

\bibitem{han}
J.~E. Han,
\newblock Phys. Rev. B 73 (2006) 125319.

\bibitem{bishop}
Z.~G. Yu, D.~L. Smith, A.~Saxena, A.~R. Bishop,
\newblock Phys. Rev. B 59 (1999) 16001.

\bibitem{renaud}
N.~Renaud, M.~Ratner, C.~Joachim,
\newblock J. Phys. Chem. B 115 (2011) 5582.

\bibitem{frederiksen}
T.~Frederiksen, M.~Paulsson, M.~Brandbyge, A.-P. Jauho,
\newblock Phys. Rev. B 75 (2007) 205413.

\bibitem{marcus}
R.~A. Marcus,
\newblock Rev. Mod. Phys. 65 (1993) 599.

\bibitem{holstein}
T.~Holstein,
\newblock Annals of Physics 281 (2000) 706.

\bibitem{ssh-rmp}
A.~J. Heeger, S.~Kivelson, J.~R. Schrieffer, W.-P. Su,
\newblock Rev. Mod. Phys. 60 (1988) 781.

\bibitem{ssh-pnas}
W.~P. Su, J.~R. Schrieffer,
\newblock Proc. Natl. Acad. Sci. USA 77 (1980) 5626.

\bibitem{ssh-prb}
W.~P. Su, J.~R. Schrieffer, A.~J. Heeger,
\newblock Phys. Rev. B 22 (1980) 2099.

\bibitem{conwell-apl}
S.~V. Rakhmanova, E.~M. Conwell,
\newblock Appl. Phys. Lett. 75 (1999) 1518.

\bibitem{conwell-synthmet2}
S.~V. Rakhmanova, E.~M. Conwell,
\newblock Synth. Met. 110 (2000) 37.

\bibitem{stafstrom-prl}
A.~Johansson, S.~Stafstr\"om,
\newblock Phys. Rev. Lett. 86 (2001) 3602.

\bibitem{fehnske}
H.~Fehnske, G.~Wellein, J.~Loos, A.~R. Bishop,
\newblock Phys. Rev. B 77 (2008) 085117.

\bibitem{tikhodeev}
S.~Tikhodeev, M.~Natario, K.~Makoshi, T.~Mii, H.~Ueba,
\newblock Surf. Sci. 493 (2001) 63.

\bibitem{ishii}
H.~Ishii, K.~Honma, N.~Kobayashi, K.~Hirose,
\newblock Phys. Rev. B 85 (2012) 245206.

\bibitem{kalosakas}
G.~Kalosakas, K.~O. Rasmussen, A.~R. Bishop,
\newblock Synth. Met. 141 (2004) 93.

\bibitem{tully}
J.~C. Tully,
\newblock Faraday Discuss. 110 (1998) 407.

\bibitem{mele}
E.~J. Mele,
\newblock Phys. Rev. B 26 (1982) 6901.

\bibitem{streitwolf}
H.~W. Streitwolf,
\newblock Phys. Rev. B 58 (1998) 14356.

\bibitem{allen}
R.~E. Allen,
\newblock Phys. Rev. B 50 (1994) 18629.

\bibitem{McKenzie}
R.~H. McKenzie, J.~W. Wilkins,
\newblock Phys. Rev. Lett. 69 (1992) 1085.

\bibitem{Su}
W.~P. Su,
\newblock Solid State Commun. 42 (1982) 497.

\bibitem{franco}
I.~Franco, P.~Brumer,
\newblock J. Chem. Phys. 136 (2012) 144501.

\bibitem{nessfisherPRL}
H.~Ness, A.~J. Fisher,
\newblock Phys. Rev. Lett. 83 (1999) 452.

\bibitem{yangkawazoe}
Y.~Yang, Y.~Kawazoe,
\newblock EPL-Europhys. Lett. 98 (2012) 66007.

\bibitem{phillpot}
S.~R. Phillpot, D.~Baeriswyl, A.~R. Bishop, P.~S. Lomdahl,
\newblock Phys. Rev. B 35 (1987) 7533.

\bibitem{Lu}
J.-T. L\"{u}, M.~Brandbyge, P.~Hedeg\aa rd, T.~N. Todorov, D.~Dundas,
\newblock Phys. Rev. B 85 (2012) 245444.

\bibitem{joachim2}
C.~Joachim,
\newblock Chem. Phys. 116 (1987) 339.

\bibitem{joachim1}
P.~Sautet, C.~Joachim,
\newblock J. Phys. C: Solid State Phys. 21 (1988) 3939.

\bibitem{geometry1}
Z.~X. Dai, X.~H. Zheng, X.~Q. Shi, Z.~Zeng,
\newblock Phys. Rev. B 72 (2005) 205408.

\bibitem{geometry2}
Y.~Xue, M.~A. Ratner,
\newblock Phys. Rev. B 68 (2003) 115407.

\end{thebibliography}
\end{document}